\documentclass[sigconf]{acmart}

\graphicspath{{figures/}}

\usepackage{algorithm}
\usepackage{algorithmic}

\usepackage{balance}

\usepackage{amssymb}
\newcommand{\pluseq}{\mathrel{+}=}
\newcommand{\minuseq}{\mathrel{-}=}

\copyrightyear{2018}
\acmYear{2018}
\setcopyright{acmlicensed}
\acmConference[MPS '18]{2nd International Workshop on Multimedia
Privacy and Security}{October 15, 2018}{Toronto, ON, Canada}
\acmBooktitle{2nd International Workshop on Multimedia Privacy and Security (MPS '18), October 15, 2018, Toronto, ON, Canada}
\acmPrice{15.00}
\acmDOI{10.1145/3267357.3267365}
\acmISBN{978-1-4503-5988-7/18/10}

\title{Lost in the Digital Wild: Hiding Information in Digital Activities}
\begin{document}

\author{Shujun~Li}
\authornote{Corresponding author: \url{http://www.hooklee.com/}.}
\orcid{0000-0001-5628-7328}
\email{S.J.Li@kent.ac.uk}
\affiliation{%
  \department{School of Computing \& Kent Interdisciplinary Research Centre in Cyber Security (KirCCS)}
  \institution{University of Kent}
  %\streetaddress{P.O. Box 1212}
  \city{Canterbury}
  \state{Kent}
  \country{UK}
  \postcode{CT2 7NF}
}
\additionalaffiliation{%
  \position{Visiting Professor}
  \institution{University of Surrey, UK}
  \department{Department of Computer Science \& Surrey Centre for Cyber Security (SCCS)}
  %\city{Guildford}
  %\state{Surrey}
  %\country{UK}
  %\postcode{GU2 7XH}
}

\author{Anthony~T.S.~Ho}
\email{a.ho@surrey.ac.uk}
\affiliation{%
  \department{Department of Computer Science \& Surrey Centre for Cyber Security (SCCS)}
  \institution{University of Surrey}
  \city{Guildford}
  \state{Surrey}
  \country{UK}
  \postcode{GU2 7XH}
}
\additionalaffiliation{%
  \position{Guest Professor}
  \institution{Wuhan University of Technology, China}
  \department{School of Computing Science and Technology}
  %\city{Wuhan}
  %\state{Hubei}
  %\country{China}
}
\additionalaffiliation{%
  \position{Tianjin Distinguished Professor \& Guest Professor}
  \institution{Tianjin University of Science and Technology, China}
  \department{School of Computing Science and Information Engineering}
  %\city{Tianjin}
  %\country{China}
}

\author{Zichi~Wang}
\email{wangzichi@shu.edu.cn}
\affiliation{%
  \department{School of Communication and Information Engineering}
  \institution{Shanghai University}
  \city{Shanghai}
  \country{China}
}

\author{Xinpeng~Zhang}
\email{xzhang@shu.edu.cn}
\affiliation{%
  \department{School of Communication and Information Engineering}
  \institution{Shanghai University}
  \city{Shanghai}
  \country{China}
}
\additionalaffiliation{%
  \institution{Fudan University, China}
  \department{School of Computing Science}
  %\city{Shanghai}
  %\country{China}
}

\renewcommand{\shortauthors}{S.~Li et al.}

\begin{abstract}
This paper presents a new general framework of information hiding, in which the hidden information is embedded into a collection of activities conducted by selected human and computer entities (e.g., a number of online accounts of one or \emph{more} online social networks) in a selected digital world. Different from other traditional schemes, where the hidden information is embedded into one or more selected or generated cover objects, in the new framework the hidden information is embedded in the fact that some \emph{particular} digital activities with some \emph{particular} attributes took place in some \emph{particular} ways in the \emph{receiver-observable} digital world.

In the new framework the concept of ``cover'' almost disappears, or one can say that now the \emph{whole} digital world selected becomes the cover. The new framework can find applications in both security (e.g., steganography) and non-security domains (e.g., gaming). For security applications we expect that the new framework calls for completely new steganalysis techniques, which are likely more complicated, less effective and less efficient than existing ones due to the need to monitor and analyze the whole digital world \emph{constantly} and in \emph{real time}. A proof-of-concept system was developed as a mobile app based on Twitter activities to demonstrate the information hiding framework works. We are developing a more hybrid system involving several online social networks.
\end{abstract}



\maketitle

\section{Introduction}

Traditional information hiding schemes embed hidden data into a cover object (e.g., a file or a communication channel) by slightly modifying its content, and hope to minimize the distortion between the original cover object and its information-bearing version \cite{Holub-Fridrich:WIFS2012, Holub:EURASIP-JIS2014, Guo:IEEETIFS2015, BinLi:IEEETIFS2015, ZichiWang:JEI2016, WeimingZhang:IEEETCSVT2017, NetworkSteganographyBook2016}. When an information hiding method is used for security purposes, it is commonly called steganography where the threat model is that the communications between the sender and the receiver are under control of untrusted third-parties like hostile authorities. Modifications made to the cover object(s) provide possibilities for steganalysis \cite{SteganalysisBook2010, Kodovsky:IEEETIFS2012, Holub-Fridrich:IEEETIFS2015, Tang:IEEETIFS2016, Xu:IEEESPL2016}. Therefore, it is necessary to invent new steganographic methods to make information hiding more secure. One new approach called generative (also known as coverless) information hiding is to move away from using a selected (pre-existing) cover object, but to embed hidden data into one or more generated objects that did not exist \cite{Otori-Kuriyama:SG2007,Otori-Kuriyama:IEEECGA2009,Wu-Wang:IEEETIP2015,Xu:VC2015,Hayes:NIPS2017,Shi:PCM2017,Volkhonskiy:NISP2018WAT,Zhang:IEEETMM2018}.

In this paper, we propose a new general information hiding framework that goes beyond the generative/coverless approach. It hides information in activities conducted by human and computer entities in a selected digital world. Examples of digital worlds include the whole Internet, the whole Web, one or more online social networks and/or web forums, an online game participated by many (human and computer) players, a Wiki website with many editors and visitors, an FTP server used by many users, a mailing list subscribed by many subscribers, or even just a shared computer used by multiple users. In this new framework, the concept of ``cover'' becomes even less relevant, as the whole digital world selected can be seen as the \emph{dynamic} ``cover'' of all past and \emph{future} hidden messages embedded. In the new framework the hidden information is represented by (i.e., embedded in) the fact that some \emph{particular} digital activities with some \emph{particular} attributes took place in some \emph{particular} ways in the \emph{receiver-observable} digital world. Here, the information may be hidden in one or more ways, e.g., meta data of some activities, a particular order and combination of some activities, selection of specific (types of) activities, and attributes of some activities' contents. The capacity of the new information hiding framework is relatively low if activities' contents are not used for hiding information, but it can easily incorporate (i.e., conditionally trigger) one or more traditional cover-based and coverless information hiding methods (e.g., image and video based methods) to increase the embedding capacity. The incorporation can take place \emph{when and only when} necessary to minimize the exposure of the high-capacity (often less secure) scheme(s) to the adversary.

For security applications, a higher undetectability can potentially be achieved using this new framework because that one can choose not to change any content of the digital activities. This is possible because the hidden information is now represented by some \emph{particular} attributes of \emph{all} the activities in the selected digital world \emph{collectively}. For non-security applications, the new framework can open up many new possibilities for embedding information in different kinds of digital worlds, e.g., hidden plots, maps, weapons, characters, and difficulty levels in an online game to increase the level of player engagement and entertainment.

Although there is quite some related work, to the best of our knowledge no any existing scheme or framework covers all the features of the proposed framework. Some ad hoc schemes proposed in the research literature can be considered special cases of the proposed framework, but they are not presented as a general framework that can be generalized to cover different types of digital activities. In Section~\ref{section:related_work} we will give a more detailed review of all related work we are aware of and how the proposed framework differs from them.

The main focus of this paper is to present the general framework as a new concept for designing information hiding schemes. Its feasibility is demonstrated via a proof-of-concept system, a mobile app allowing transmission of hidden messages between two persons using Twitter as the digital world. The capacity allows short messages exchanged between two persons in a rate comparable to SMS messages over mobile phones. The mobile app is designed more for fun (non-security purposes), but it can be extended towards a more secure system. We are currently implementing a larger proof-of-concept system involving more than one online social networks and multiple types of activities, for security applications.

The rest of the paper is organized as follows. We introduce related work in Section~\ref{section:related_work}. Section~\ref{section:Architecture} describes the architecture and general steps of the proposed framework. Then, three core algorithms in the proposed framework are introduced in Section~\ref{section:Algorithms}. After the core algorithms are explained, Section~\ref{section:Demonstrator} describes the proof-of-concept system we implemented to demonstrate the feasibility of the proposed framework. Section~\ref{section:Extensions} lists some possible extensions to the basic framework. The last section concludes the paper.

\section{Related Work}
\label{section:related_work}

The idea of hiding information in activities has been used in other contexts such as games (e.g., the bidding system in contract bridge card games) and intelligence (e.g., secret agents reportedly used physical activities to encode simple secret messages). There have been some attempts of using activities in games to create covert channels, but such work is all very ad hoc and cannot be directly generalized to other digital activities. There has been some related work on information hiding with online social networks, which can be seen as special cases of the proposed framework. Some work in network steganography may be considered loosely related as well because most methods used meta data to hide information, which may be argued as special digital activities. In this section, we systematically categorize selected work on all types of information hiding methods that are related to the proposed new framework and explain how the new framework differs from them.

\subsection{Selected Cover(s)}

Covers used in information hiding are normally not fixed, so the sender can freely decide what cover(s) to use for hiding a specific message. There has been quite some work on how to better select covers for increasing security and/or capacity. For instance, in \cite{Kharrazi:ICIP2006}, images with larger changeable DCT coefficients and higher quality factor are regarded as suitable images for secret data carrying. For uncompressed image, suitable images are selected according to image texture and complexity \cite{Sajedi:CIT2008}\cite{Sajedi:IJIS2010}. The images with more complex texture are preferentially selected. A unified measure to evaluate the hiding ability of a cover image is proposed in \cite{Wu:ICIMCS2015} by representing images using the Gaussian mixture model and formulated the measure in terms of the Fisher information matrix.

Our proposed new framework may involve selection of activities that can be better used for information hiding purposes, but the contents of those selected activities are not normally used as the cover for hiding information.

\subsection{Batch Steganography}

Batch steganography aims to hide information securely by spreading across a batch of covers, and the dual problem is called pooled steganalysis -- to obtain more reliable detection of steganography in large sets of objects \cite{Ker:IH2006}. With batch steganography, the security performance is improved with the increasing of the quantity of covers \cite{Ker:SPL2007}. In \cite{Ker:MMSec2012}, five payload-allocation strategies were proposed for different scenes. In these strategies, the sender must know the maximum length of secret data that he can embed into each given image with the chosen steganographic method. So, the strategies can only use for early non-adaptive steganographic methods. A payload-allocation strategy for modern steganography was proposed in \cite{Zhao:IWDW2016}. The payload to be allotted for each image is calculated by equating the steganographic distortion of all the images.

Our proposed new framework has some flavor of batch steganography since normally multiple activities are involved for embedding the hidden information. We however go beyond batch steganography as relationships between selected activities and how they appear in the whole digital world can also be used to embed information. In addition, batch steganography still uses the content of each cover for information hiding, which is normally not the case for our framework.

\subsection{Generative (Coverless) Information Hiding}

The idea of generative information hiding probably appeared firstly in 2007 \cite{Otori-Kuriyama:SG2007,Otori-Kuriyama:IEEECGA2009}, when Otori and Kuriyama proposed to embed information by synthesizing texture images with different repetitive texture patterns. Otori and Kuriyama's scheme was improved by Wu and Wang in \cite{Wu-Wang:IEEETIP2015}, but the new scheme was shown insecure by Zhou et al.\ \cite{Zhou:IEEETIP2017}. In \cite{Xu:VC2015} a different scheme of generating marbling textures to deliver hidden information. More recently, machine learning based approaches have been considered to generate covers for steganography \cite{Hayes:NIPS2017,Shi:PCM2017,Volkhonskiy:NISP2018WAT}. In \cite{Zhang:IEEETMM2018}, Zhang et al.\ proposed a coverless steganography method based on Discrete Cosine Transform (DCT) and Latent Dirichlet Allocation (LDA) topic classification, where images that each represent a segment of the hidden message are selected from a database.

Our proposed new framework may involve generated activities for hiding information, but it can also work with modified meta data or even content of selected activities so it supports hybrid hiding methods.

\subsection{Network Steganography}

Network steganography techniques use network traffic as a carrier for the secret data \cite{NetworkSteganographyBook2016}. The various characteristic features such as control information, behavior, or relationships of network protocols can be utilized to conceal data in network traffic. These techniques can be classified into storage and timing methods based on how the secret data are encoded into the carrier. A real-time inter-protocol steganographic method was proposed in \cite{Lehner:LCN2017}. It utilizes relationships between two or more overt protocols to hide data, such as the relationship between Real-Time Transport Protocol and Real-time Control Protocol. Meanwhile, a data hiding detection method which relies on network traffic coloring is introduced. In \cite{Moodi:JCS2018}, the least significant bits algorithm and an algorithm based on delayed packets in VoIP streams are combined to forms a new algorithm. The probability of detection existing steganography data is reduced extensively using the combined algorithm.

The proposed new framework could involve activities taking place at the network protocol level, but in most cases it will work at the application and/or user level(s), i.e., it will mostly work with activities at the highest layer of the network stack. Some ideas used in network-based steganography can be used in the proposed framework, e.g., the timing information of some specific posts from some target accounts can be used to encode information in a way similar to how timing information is used at the lower network protocol layer.

\subsection{Game-based Information Hiding}

This approach conceals information in games, typically by hiding information in game elements such as the chess board or sequence of items generated \cite{Naomoto:STEG2002,Lee:ICMLC2008,Ou:IS2014} or how game players behave during the game \cite{Murdoch-Zielinski:GameSteganography:IH2004,Julio:CS2006,Desoky:SCN2009,Ritchey:XRDS2013}. Such information hiding methods has its root in some popular games with multiple players who group into teams, contract bridge being a well-known example in which players in the same team use various activities based approaches to exchange secret messages between them during the biding phase \cite{Bridge2018}.

Some game-based information hiding schemes can be seen as special cases of the proposed new framework, where the game is the selected digital word and the information-bearing activities are game players' behaviors during one or more game sessions. None of the game-based information hiding schemes have been designed to allow further generalization to other digital worlds and activities.

\subsection{Social Network Based Information Hiding}

A phenomenon called social steganography was described in \cite{Knudsen:SM2013}, where hidden messages can be exchanged in social network. For example, regular intervals posted messages on Facebook that on the surface seemed innocent updates, but a hidden message can be included. In \cite{Pantic:ACSAC2015}, a steganographic system based on the online social network platform Twitter was proposed, where the secret messages are embedded into the length of tweets. In \cite{Gilbert:CHI2015}, an algorithm was proposed to cloak messages by transforming them to resemble the average Internet messages. A steganographic method converting the secret message into ``love'' marks in WeChat (a popular instant messaging system in China) is proposed in \cite{Zhang:IH-MMSP2016}. The sender can let the ``love'' marking rate on each friend be a normal given value to keep the covert communication confidential, and the receiver can extract the secret message from only the visible ``love'' marks.

Most social network based information hiding schemes can be seen as special cases of the proposed new framework, but what has been reported in the literature is mostly ad hoc methods. Work in this category does not go beyond online social networks like the proposed framework does.

\subsection{Other Behavioral Based Information Hiding}

Behavior based information hiding can be traced back to the pre-WW2 and cold war ages when security agents often used specific activities or behavior to deliver secret messages \cite{Danigelis:Seeker2012}. For instance, one method for covert communication was how shoelaces were tied: connecting them between the holes on both sides of a shoe in different ways signaled certain things such as ``follow me'' or ``I have brought another person''.
%Hollow coins are traditional information hiding based on objects, not based on behaviors or activities.
%Furthermore, hollow coins could carry messages. Although the space inside was extremely small, agents could put in a microdot. This micro writing system developed by the CIA in the 1960s and 70s required a high-powered magnifier to read concealed messages.

The proposed framework may be understood as extending some ideas security agents used in the physical world to the cyber world. This move allows a high level or full automation of the information encoding and decoding processes, and opens up the possibilities of using a lot of new types of activities and digital worlds.

\begin{figure*}[!t]
\centering
\includegraphics[width=0.6\linewidth]{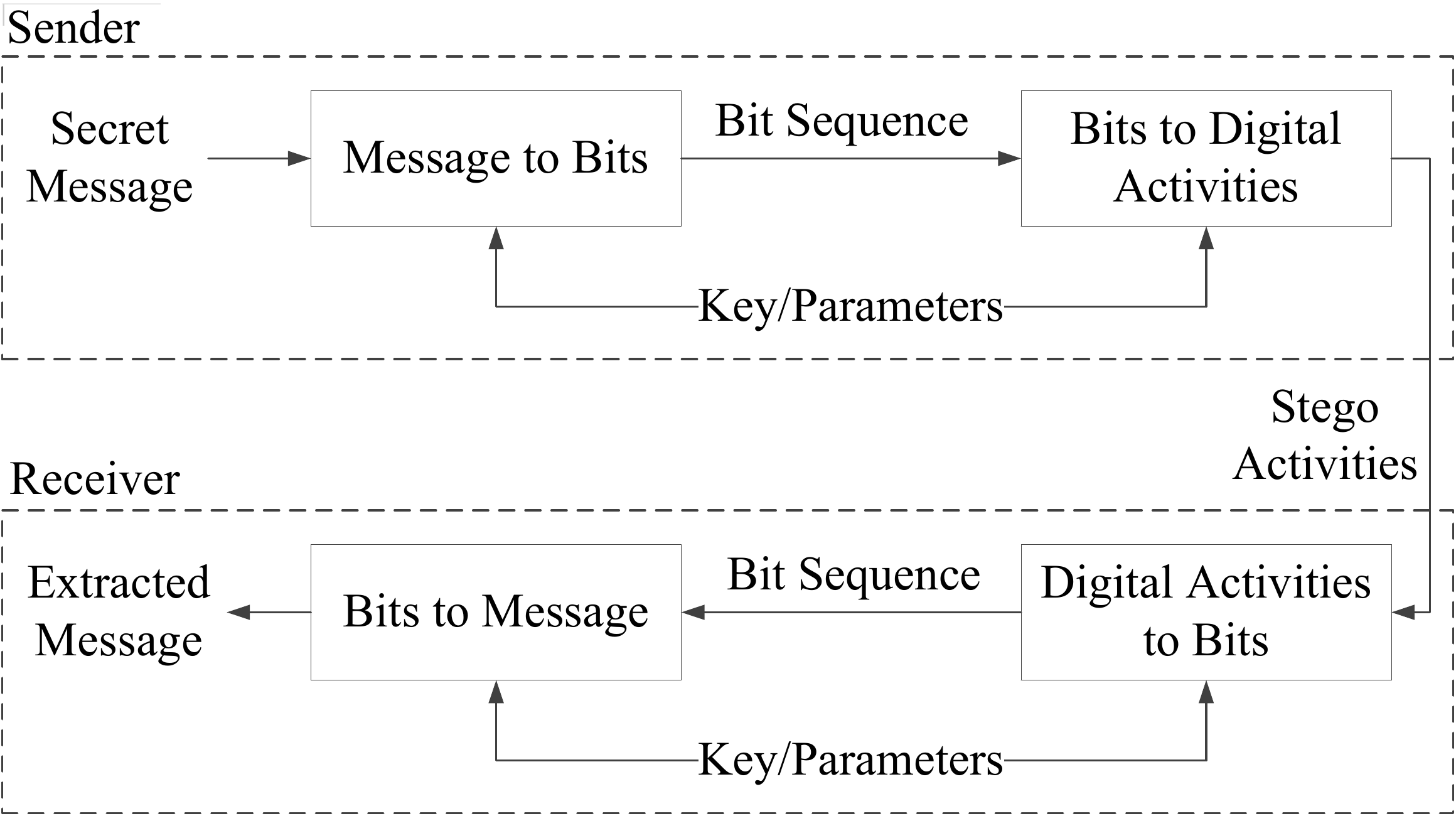}
\caption{Overall architecture of the proposed information hiding framework.}\label{fig:architecture}
\end{figure*}

\section{Overall Architecture}
\label{section:Architecture}

In this section, we explain the basic ideas behind the proposed framework, a high-level abstraction of the encoding and decoding processes, potential applications, and how the activities in the framework can be implemented in real world.

\subsection{Basic Ideas}

The proposed framework represents the secret message as activities in a selected digital world. Here, the digital world can be anything where the sender and the receiver can communicate, e.g., a computer network such as the Internet and the Web, or just a computer with multi-user support. The secret message is not normally embedded in the contents of activities as in traditional methods, but represented as the fact some \emph{particular} digital activities with some \emph{particular} attributes took place in some \emph{particular} ways in the \emph{receiver-observable} digital world. The ``cover'' in the traditional sense can be seen as the whole digital world. Since the contents of information-bearing activities are not modified during the embedding process, it has the potential to become more difficult to detect for increased security. The overall architecture of the proposed framework is shown in Figure~\ref{fig:architecture}.

The proposed framework partly addresses Open Problem 11 described by Ker et al.\ in 2013 \cite{Ker:IHMMSec2013}: ``Technical and societal aspects of inducing randomness in communications to simplify steganography.'' In the context of this open problem, Ker et al.\ discussed how ``to engineer the real world so that parts of it match the assumptions needed for security proofs.'' To some extent, the proposed framework gives a practical approach to engineering the real world by embedding hidden information in the set of real-world activities. The proposed framework also touches three other open problems described in \cite{Ker:IHMMSec2013}: i) Open Problem 7 ``Theoretical approaches and practical implementations for embedding in multiple objects in the presence of realtime constraints.'' ii) Open Problem 9 ``How to perform cover selection, if at all? How to detect cover selection?'' iii) Open Problem 19 ``Any detector for multiple objects, or based on sequential hypothesis tests.'' Note that Ker et al.'s paper focuses on steganography (security applications) only, but the proposed framework covers both security and non-security applications.

\subsection{General Encoding and Decoding Processes}

The encoding process can be largely split into three steps:
\begin{itemize}
\item
\textbf{Step 1}: Converting the message $M$ into a bit sequence $\{b_i\}$.

\item
\textbf{Step 2}: Converting $\{b_i\}$ into a number of activities $\{A_i\}$.

\item
\textbf{Step 3}: Using one or more entities to conduct $\{A_i\}$.
\end{itemize}

Once a hidden message is sent to the digital world, the receiver can run the following process to detect and decode the message:
\begin{itemize}
\item
\textbf{Step 1}: Scanning all relevant activities of the target entities' $\{A\}$ to find out if some activities represent a hidden message.

\item
\textbf{Step 2}: Once a subset of information-bearing activities $\{A_i\}$ are located, converting the activities into a bit sequence $\{b_i\}$.

\item
\textbf{Step 3}: Converting $\{b_i\}$ to a message $M'$ which is readable to the intended receiver (depending on if the message is for a human receiver to read or a computer program to do some automated work).

\item
\textbf{Step 4}: Checking the decoded message's integrity and go back to Step~1 if the message does not pass the integrity check. This step may be done as part of Step(s)~2 and/or~3.
\end{itemize}

The whole processes of sending and receiving a hidden message can be controlled by a secret ``key'' and/or a number of public parameters which can influence different steps of the processes.

\subsection{Applications and Implementations}

As most other information hiding methods, the proposed framework can be used in not only security applications but also non-security applications. In security applications, there will be a secret key that may be used to determine some or all of the parameters. In non-security applications, a key is normally not needed so the whole process is essentially public. For security applications, we also assume the sender and the receiver both possess a trusted device that runs a trusted computer program implementing the proposed information hiding framework.

Typical implementations of digital activities are those done by online accounts registered with websites and Internet services. However, many other online activities that may happen over the Internet or any other types of computer networks can also be used as long as i) the entities involved can be controlled by the sender to conduct the activities and ii) the receiver can read the activities and link them with the entities. To simplify our discussion in the remaining parts of this paper, we will focus on online activities conducted by Internet/web user accounts.

\section{Core Algorithms}
\label{section:Algorithms}

There are three core algorithms in the proposed framework: encoding algorithm, decoding algorithm, and an algorithm for avoiding conflicts of simultaneous access of the hidden communication channel by multiple senders (which is necessary for the core algorithms to be robust and efficient in real world).

For security applications, the security of the core algorithms depends on the ``naturalness'' of the online activities used to carry a hidden message, where ``naturalness'' is defined as to what extent the information-carrying activities are indistinguishable from other hidden information free activities. Note that the information-carrying activities may not be only those generated by human users, so the ``naturalness'' is not only about natural human behavior. Since we are proposing a general framework, we leave security analysis of different implementations of the framework as future work. Instead, our focus here is on the design of the core algorithms and the development of a proof-of-concept system of the basic framework.

To ease the descriptions of the core algorithms, we make a number of assumptions and define some notations. We assume that the hidden message to be sent has been converted to a bit sequence $M$. The encoding and decoding algorithms are based on $n$ ``bits to activities'' mappings denoted by $\text{B2A}_1(\cdot)$, ..., $\text{B2A}_n(\cdot)$. Each $\text{B2A}_i(\cdot)$ can map $B(i)$ bits into $A(i)$ digital activities conducted by one or more entities. A key $K$ may be used to control all the ``bit to activities'' mappings and other steps of the core algorithms when the algorithms are applied to security applications. Note that the ``bit to activities'' mappings can also be selected from a large set of candidate mappings under the control of the key $K$.

\subsection{Encoding Algorithm}

The encoding algorithm is a two-step process as follows.
\begin{itemize}
\item
\textbf{Step 1} (optional): Expand $M$ by adding $s\geq1$ markers $m(1)$, ... , $m(s)$. Denote the expanded message by $M^*$. The markers may be added to any positions of $M$, and the positions are derived from either $K$ or a number of public parameters.
Some situations need to add markers are as follows:
\begin{itemize}
\item
Some markers may be dependent on $M$, e.g., a hash value of $M$ for integrity checking purposes.

\item
Some markers can be used to ease localization of the hidden message in digital activities at the decoder side. It is optional because there are other ways to localize the hidden message without using a marker (e.g., measuring naturalness of decoded candidate messages).

\item
One marker may be added to the beginning of $M$ (``start marker''), signaling the start of a message for streamlining the decoding process.

\item
One marker may be appended to the end of $M$ (``end marker''), signaling the end of a message for error control purposes and for streamlining the decoding process.
\end{itemize}

\item
\textbf{Step 2}: Convert the expanded message $M^*$ into a number of digital activities $A(M^*)$. The process is better represented by the following pseudo code in Algorithm~\ref{algorithm:B2A}.
\end{itemize}

\begin{algorithm}
\caption{Bits to Activities Mapping}
\label{algorithm:B2A}
\begin{algorithmic}
\STATE $A(M^*)=\varnothing$
\REPEAT
	\FOR{$i=1$ to $n$}
		\IF{there are less than $B(i)$ bit in $M^*$ to be encoded}
		\STATE pad $M^*$ by a pre-defined bit pattern to have $B(i)$ bits for further encoding, where the padding bit sequence can be static or dynamically generated
		\ENDIF
		\STATE $M^*_i=$ the next $B(i)$ bit of $M^*$
		\STATE $A(M^*) \pluseq \text{B2A}_i(M^*_i)$
		\STATE $M^* \minuseq M^*_i$
		\IF{$M^*=\varnothing$}
			\STATE \textbf{break}
		\ENDIF
	\ENDFOR
\UNTIL{$M^*=\varnothing$}
\end{algorithmic}
\end{algorithm}

Each activity in $A(M^*)$ also contains needed (absolute and /or relative) timing information about when the activity should be conducted by the corresponding information-carrying account.

Any ``bits to activities'' mapping $\text{B2A}_i(\cdot)$ can be used in the above encoding algorithm as long as they are invertible, meaning there is an inverse mapping $\text{A2B}_i(\cdot)$ which fulfills $\text{A2B}_i(\text{B2A}_i(x)) = x$ for any given bit sequence $x$ of size $B(i)$. A mapping may involve one or more entities (accounts) and one or more sites/services. The main requirements of a ``good'' mapping include at least the following:
\begin{itemize}
\item
\emph{Functionality}: The mapping should do the work as expected to convert a bit sequence to a finite number of digital activities that can be reversed back to the original bit sequence.

\item
\emph{Capacity}: The mapping is expected to produce a reasonable information hiding capacity. Note that the capacity is measure as the number of hidden bits transmitted per second.

\item
\emph{Security} (for security applications only): The mapping should not introduce unusual patterns in the generated activities in the selected digital world so that detection of the presence of hidden messages is possible.

\item
\emph{Usability}: In case the mapping requires involvement of human users (e.g., to solve a CAPTCHA for posting a message), it should be easy and not time consuming to conduct any required manual operations.

\item
\emph{Robustness}: The mapping should be robust to unintended errors in the generated digital activities (which can either be caused by the service providers or network failures). This is the least important feature and may not be considered in error-free environments.
\end{itemize}

When the selected digital world is the Internet, some example mappings are the following:
\begin{itemize}
\item
Posting messages and/or status updates on a website (e.g., a social networking website or a web forum) or an Internet service (e.g., an instant messaging service).

\item
Replying, quoting, commenting, re-posting a message and/or status update on a website or an Internet service.

\item
Private messages sent on a web forum to one or more user accounts. In this case the receiver of the hidden message needs to have access to one of the receiving user accounts.

\item
Following or un-following another user account.

\item
The fact of uploading a particular file to a particular website at a particular time.

\item
Visiting a website or an FTP server using a particular manner (e.g. browsing different web pages of a website to form a particular order and browsing each page for a particular amount of time). In this case the receiver of the hidden message needs to have access to the visit log of the website or the FTP server, i.e., he/she is an administrator.
\end{itemize}

In the next section, we will give more details about two mappings implemented in the proof-of-concept system.

\subsection{Decoding Algorithm}

The decoding algorithm can be roughly split into four steps:
\begin{itemize}
\item
\textbf{Step 1} (only when start markers are used in the encoding process): From a given time $t_0$ scanning for the first start marker. The starting time of the scanning process can be defined by the user or automatically set as the last time when a hidden message was successfully decoded.

\item
\textbf{Step 2}: Locating the start of the expanded message $M^*$ and decoding it.

\item
\textbf{Step 3} (only when markers are used in the encoding process): Verifying all markers (which may include a padding pattern at the end) in the decoded message and remove them from $M^*$ to get the original hidden message $M$.

\item
\textbf{Step 4} (optional): Checking ``natural'' (e.g., linguistic and statistic) characteristics to see if the decoded message is a real hidden message. This step can be done manually by the receiver, but when there are too many false positives (e.g., no error control markers are added and the error rate is high) it will be important to let the decoder (a computer program) to automatically detect some (if not all) false positive messages.
\end{itemize}

The actual decoding process is more involved because Step~1 depends on a partial decoding process so it is mixed with Step~2. In addition, in case no any markers are inserted, Step~4 is better incorporated into Step~2 to make a more effective use of memory, otherwise one needs to decode a long message and then look for possible hidden message(s). From the above issues, we represent the whole decoding process using the pseudo code in Algorithm~\ref{algorithm:decoder}.

\begin{algorithm}
\caption{Decoding Algorithm}
\label{algorithm:decoder}
\begin{algorithmic}
\IF{the system is configured to have a start marker at the beginning of $M^*$ in the encoding process}
	\REPEAT
		\STATE search for the first start marker in $A$ from $t_0$
	\UNTIL{a start marker is detected}
	\STATE set $t_0$ to be the time immediately after the first start mark appears
\ENDIF
\STATE $A =$ {all activities of all information-carrying accounts from a given time $t_0$ to the current time}
\STATE $M = $ "" (empty string)
\STATE stop flag = false;
\REPEAT
	\FOR{$i = 1$ to $n$}
		\IF{there are less than $A(i)$ activities in $A$}
			\STATE stop flag = true
			\STATE \textbf{break}
		\ENDIF
		\STATE $A_i =$ the sequence of the next A(i) activities that have not been processed
		\STATE $M \pluseq \text{A2B}_i(A_i)$
		\STATE $A \minuseq A_i$
		\IF{the system is configured to have an end marker at the end of $M^*$ in the encoding process}
			\IF{the newly added part of $M$ contains an end mark}
				\STATE remove the end marker and any bits after it from $M$
				\STATE stop flag = true
				\STATE \textbf{break}
			\ENDIF
		\ENDIF
	\ENDFOR
\UNTIL{$A=\varnothing$ or the stop flag is true}
\IF{the decoder is configured to check ``naturalness'' of M in real time}
	\IF{$M$ does not pass the naturalness check}
		\RETURN NULL
	\ENDIF
\ENDIF
\RETURN $M$
\end{algorithmic}
\end{algorithm}

While the above algorithm defines a quite clear process, the main purpose of the pseudo code is to show what the decoder is supposed to do rather than how the algorithm should be implemented. Alternative ways of implementing the same algorithm is possible and encouraged for optimizing the use of memory and time. The naturalness checking step in the above algorithm is not defined so it is a sub-algorithm that can be studied separately. When synchronization markers are used, however, the naturalness checking is usually not needed. Even when synchronization markers are not used, it is still possible to depend on human users to decode hidden messages as long as the decoded message is not too long. Therefore, we do not consider the naturalness checking sub-algorithm as a very important component for the basic framework.

\subsection{Conflict Avoidance Algorithm}

The above encoding algorithm will take time to encode a hidden message. Communicating the activities to the network using the information-carrying accounts will normally take an even longer time since all the activities need posting to the network reliably. Depending on the ``bits to activities'' mappings, the timing order of activities may not matter. It is also possible to use multiple threads to conduct those activities simultaneously if there are multiple network interfaces. However, in most practical cases, we need a way to serialize all the activities so that they are posted to the network in a deterministic order since the decoding process looks at activities sequentially.

Since encoding and sending a hidden message take time, conflicts may occur if multiple senders sharing the same hidden communication channel want to send hidden messages at the same time (or at different times but the duration of the encoding/sending processes overlap). This may happen when the two parties of a communication channel wants to send hidden message simultaneously or when there are multiple senders who want to send hidden messages to the same receiver using the same channel. Both cases can be avoided by allocating different communication channels (i.e., different sets of information-carrying accounts) to 1) different directions of communications between two parties; 2) any different pairs of communicating parties. Such a simplistic solution suffers from the insufficient use of system resources and inconvenience users as more keys need to be managed by each user.

The problem can be solved by a send-and-check process at the encoder side. Such a process can be designed following similar protocols in other areas, e.g., telecommunications and operating systems.

\section{Proof-of-Concept System}
\label{section:Demonstrator}

This section describes a proof-of-concept system of the proposed information hiding framework as a demonstrator and a preliminary evaluation of its usability through a lab-based user study.

\subsection{Design and Implementation}

In the proof-of-concept system, the online social network platform Twitter is selected as the digital world and selected Twitter accounts under control of the sender are used as the information-carrying accounts. Two ``bits to activities'' mappings were designed. One mapping is based on the retweeting functionality: given $2^d$ information-carrying accounts each is associated with $2^e$ target retweeting accounts, the mapping converts a $(d+e)$-bit sequence into a unique pair of (information-carrying account, retweeted account). The other mapping is simpler and only encodes one bit: it selects one of two ways to retweet the selected retweeted account's tweet (the two ways are: direct retweeting and tweeting by simply quoting the retweeted account's tweet). Note that the second mapping does not generate any separate activity but modifies the activities generated by the first mapping, which is a way how two mappings can be combined to form a more complicated mapping. In total the two mappings give a capacity of $(d+e+1)$ per activity. The actual values of $d$ and $e$ in this proof-of-concept system are $d=5$ and $e=8$. The embedding capacity depends on the number of activities posted on Twitter, but our experiments showed it can reach the level of sending a typical short message in just one or two minutes, so comparable with SMS messages on mobile phones (which can have a delay up to a few minutes).

\begin{figure}[!htb]
\centering
\includegraphics[width=0.49\linewidth]{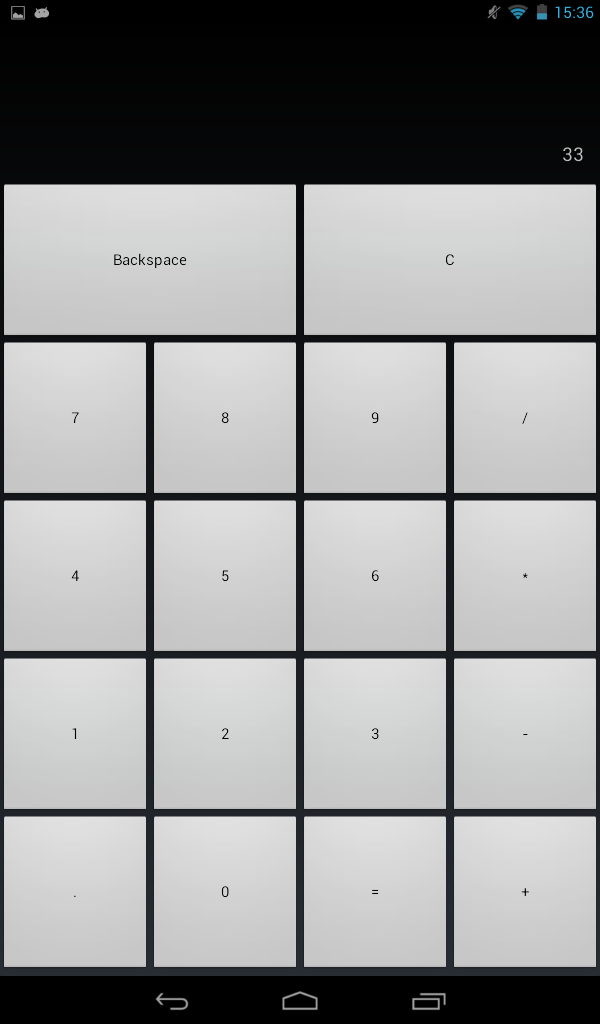}
\caption{Snapshot of the demonstrated mobile app - camouflage functionality.}\label{fig:PoC_camouflage}
\end{figure}

For the proof-of-concept system we decided to include two markers: one start marker and one end marker of size 14 bits (dynamically generated from a key) to signal the start and the end of the hidden message, respectively. The hidden message is encrypted by AES (128-bit key, PKCS5 padding, ECB mode to minimize error propagation) before being encoded. The end of the hidden message is padded using a random bit sequence generated by the key. Twitter API is used to automate the process of posting retweets and checking the status of control accounts.

The proof-of-concept system is implemented as a mobile app running from an Android device. The ``camouflage'' functionality of the mobile app is a simple calculator and the magic gesture for activating the information hiding functionality is ``pressing the <C> button for seven times''. This magic gesture is purely indicative and more complicated ones could be easily implemented. Figure~\ref{fig:PoC_camouflage} shows a snapshot of the camouflage interface of the mobile app.

After the magic gesture is given, the mobile app will show an interface with three buttons ``Encode'', ``Decode'' and ``Settings''. The first two buttons allow sending and receiving hidden messages and the last one allows viewing control accounts used and checking their online status.

\begin{figure}[!htb]
\centering
\includegraphics[width=0.98\linewidth]{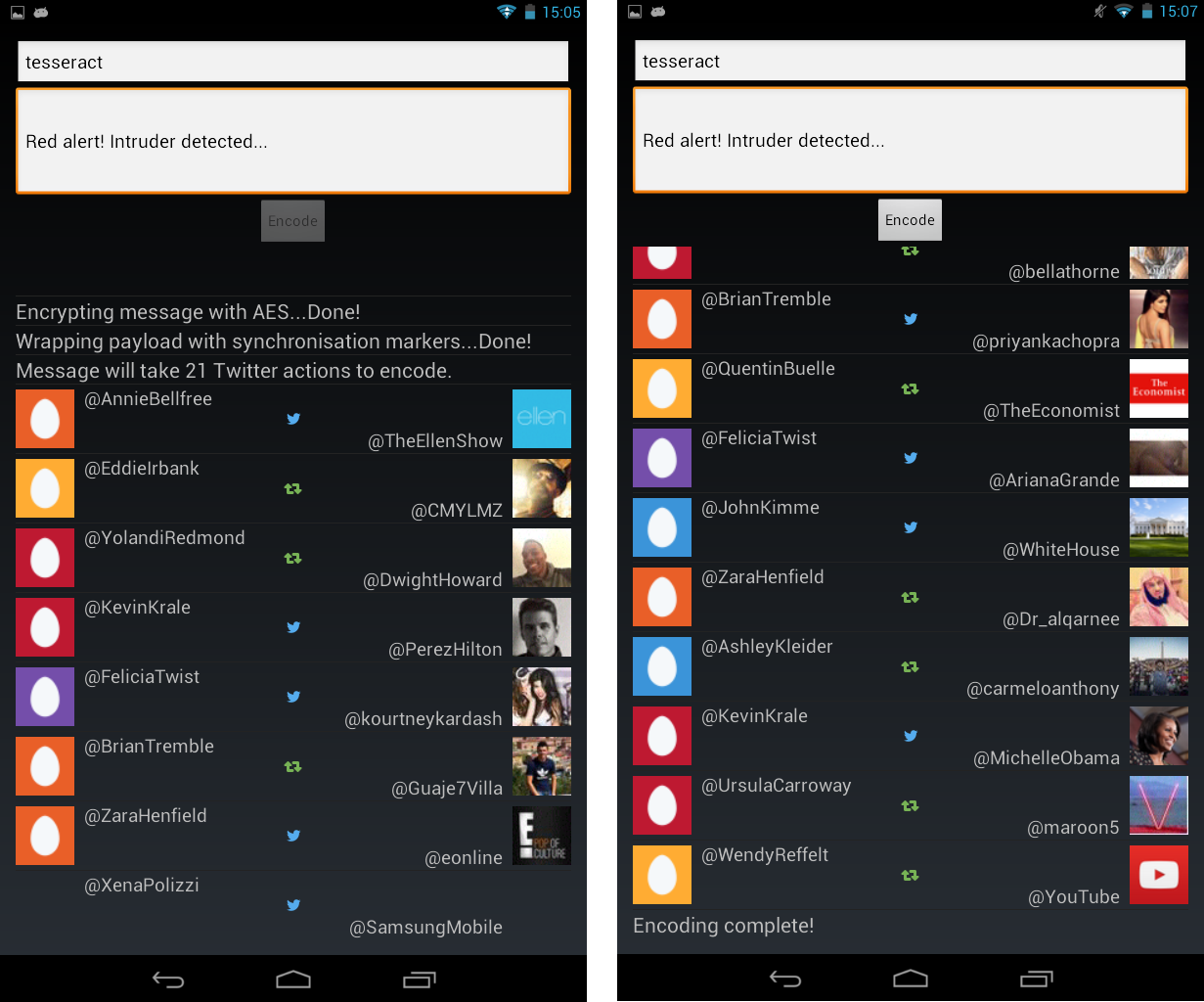}
\caption{Snapshot of the demonstrated mobile app - encoding a hidden message.}\label{fig:PoC_encoding}
\end{figure}

\begin{figure}[!htb]
\centering
\includegraphics[width=0.98\linewidth]{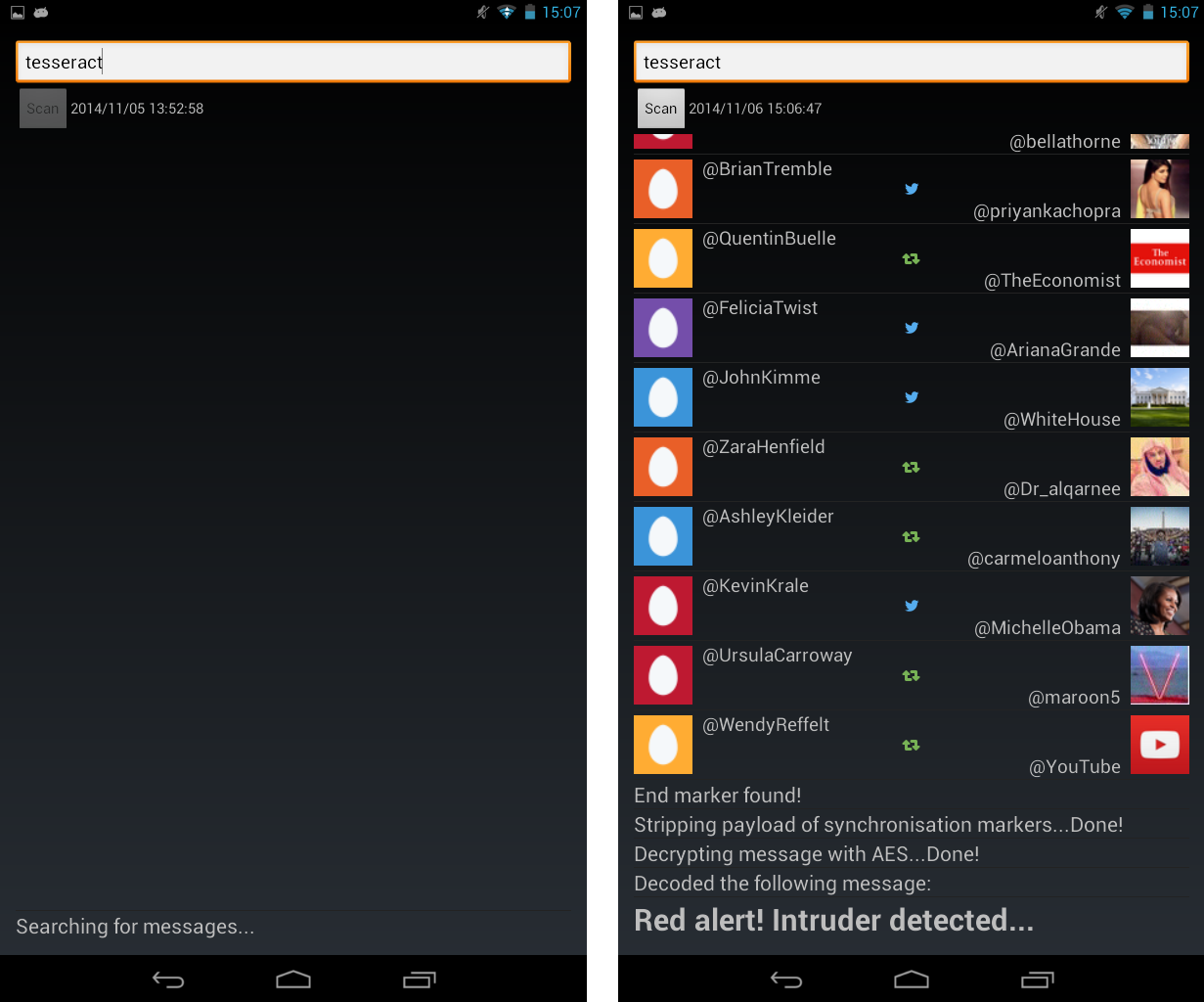}
\caption{Snapshot of the demonstrated mobile app - scanning for (decoding) a hidden message.}\label{fig:PoC_scanning}
\end{figure}

Click the ``Encode'' button, one can enter a key and a hidden message and then click the ``Encode'' button to send the hidden message to Twitter. Figure~\ref{fig:PoC_encoding} shows two snapshots of the encoding process of an example hidden message. The actual speed of encoding a message of 16 bytes is roughly 1 minute.

Click the ``Decode'' button, one can enter a key and then click the ``Scan'' button to retrieve any possible hidden messages sent previously using the same mobile app and the same key. Figure~\ref{fig:PoC_scanning} shows two snapshots of the scanning (decoding) process of an example hidden message.

The mobile app can be downloaded from \url{http://www.hooklee.com/Papers/Data/MPS2018/MobileMagicMirror.zip}. To test it two persons need to both install the app and agree on a key. The mobile app is not equipped with a conflict avoidance algorithm and by default uses some fixed Twitter accounts (created for the proof-of-concept system), so if more than two persons are using it at the same time, the communications may not be successful. We may develop an updated version of the mobile app with a basic conflict avoidance algorithm in future.

\subsection{Security}

As a first proof-of-concept system, the mobile app is designed to show wider (security and non-security) applications of the proposed information hiding framework, so making a highly secure system is not our main focus. As a matter of fact, the activities of information-carrying accounts used in the proof-of-concept system contain only retweets of a fixed $2^e$ target accounts. Such an obvious behavioral pattern can be easily detected by both humans and machines. In 2017 we conducted an experiment with human participants on detecting such accounts \cite{Li:CYBCONF2017b}, and it did show that most human participants had noticed the unusual retweeting-only behavior. While detecting an information-carrying account used in the proof-of-concept system may be relatively easy, the distribution of the hidden information across $2^d$ accounts means that the probability for an adversary to correctly detect all the information-carrying accounts and recovering the right order for conducting an analysis of the whole hidden message will be low as long as $d$ is reasonably large. Detecting any single account among all accounts in Twitter is also a hard task if the adversary has no previously known target accounts to focus its investigation on.

At the system level, there is also a demand of hiding the information hiding functionality itself. This is important when an untrusted person or organization (e.g., a hostile governmental body) gains access to the trusted device. When the hostile governmental body is involved, the issue has to be considered in the context of digital forensics where more advanced software and hardware tools can be used to scan the whole device. This new demand requires no long-term sensitive data being stored in volatile or non-volatile memory of the trusted devices. Instead, such information should be kept for a minimum time, e.g., for displaying a decoded short message on the screen for just 10-20 seconds. In addition, this challenge requires information hiding at the system level, i.e., hiding the information hiding functionality itself. This means that the information hiding functionality should be hidden from the visible user interface and is triggered only by some secret activation signal.

As to the ability in resisting existing steganalytic methods, the proposed framework has the potential to work well due to the disappearance of the ``cover''. Most modern steganalytic tools use supervised machine learning to detect signs of steganography by investigating the models of covers and stego-objects. Features are extracted from a set of labeled objects to train a common steganalytic model, and then used to distinguish a suspicious object. In the proposed framework, the feature extraction operation cannot be carried on easily since the stego objects are effectively the whole digital space. In pooled steganalysis, it is assumed that the warden already possesses a quantitative detector for whatever type of steganography the sender is using, an estimator for the length of the hidden message in an individual stego object as a proportion of the maximum length. This is an assumption hardly true for the proposed framework.

To improve security, some extensions to the basic framework can be considered, some of which are discussed in Section~\ref{section:Extensions}.

\subsection{Usability Evaluation}

Although our own testing of the mobile app showed that it was very easy to use, we wanted to have a more proper evaluation of its usability from normal users' perspective. To this end, we conducted a usability evaluation experiment. In the evaluation, in total 15 participants were recruited, including 11 males and 4 females with different demographic backgrounds.

On a scale of 5 points, the user rated encoding and decoding times. The median ratings for both times are 3, and the averages are 2.87 and 3.20, respectively. It seems participants tended to have more concerns on the relatively slow encoding time.

We also asked all participants what their maximum tolerances for the encoding and decoding times were, and the answers vary a lot (ranging from 1 second to 2 minutes for the encoder and from 1 second to 1.5 minutes for the decoder). The median tolerance for the encoding time is 30 seconds and that for the decoding time is halved. The average tolerances for the encoding and decoding times are 30.50 and 27.67, respectively, but we consider these average values are less informative due to the high variances of user ratings.

The most interesting results are those about the likelihood to use the mobile app in real world. One participant gave the lowest rating (1) and three participants gave the highest rating (5). The median rating is 3 and the average is 3.47.

Looking at the results as a whole, we felt that the proof-of-concept system is largely acceptable for exchanges of basic textual messages. Note that the embedding capacity can be easily increased by incorporating a high-capacity information hiding method into the proposed framework (see Section~\ref{section:CascadedIH} for more details).

\section{Possible Extensions}
\label{section:Extensions}

What we described so far is just the general framework and a relatively simplistic proof-of-concept system. The framework can actually be enhanced by some extensions in order to improve its functionalities and the overall performance. Some of such extensions can contribute to a higher level of security by making the information-carrying activities more ``natural'' and/or reducing the exposure of less ``natural'' activities. In the following, we briefly describe some selected extensions that we have considered.

\subsection{Partial Embedding: Decoy Activities}

The ``bits to activities'' mappings used in the encoding algorithm do not have to use all activities generated to carry the hidden information. Instead, a ``bits to activities'' mapping can also generate ``decoy'' activities that to make detection of hidden information even harder. Such decoy activities can be linked with those information-carrying activities so that the activities of each information-carrying account look more natural. For instance, if one information-carrying activity is about reposting a particular user's tweet, then a decoy activity can be generated to repost another related tweet of the same user.

We call the technique of not using the whole capacity of a covert channel for information hiding purpose ``partial embedding'', following a similar concept in traditional information hiding methods.

A partial embedding mapping can be designed by defining a way to generate decoy activities and for decoding the hidden information the inverse mapping can simply ignore all such decoy activities. One naive example is to simply generate a double number of activities and use the odd-numbered activities for information hiding purpose. A slightly more advanced approach is to generate a pseudo-random bit sequence and then use it to decide which activities are decoys.

The above partial embedding idea can be further enhanced by involving the human user (the sender) to define the contents of the generated decoy activities. This can help improve the security of the information hiding framework. This can be done by two approaches: 1) asking the human user to fill the contents in the encoding process; 2) collecting contents of real activities conducted by the human user in advance (before the encoding process starts) and then automatically filling the generated decoy activities. The first approach will require the human user's active involvement in the encoding process and can lower usability and prolong the whole encoding process. The second approach is a more desired solution, but requires active collection of the human user's online activities and a ``smart'' algorithm to decide what real online activities should be used for what decoy activities.

\subsection{Modulating Real Online Activities}

Rather than using real online activities for decoy activities, it will be even more secure if we use real online activities for information hiding purpose as well. In other words, we may be able to modulate real online activities in some way to convey additional hidden information without touching the semantic contents of the real online activities. This requires a queuing mechanism where human users' real online activities and hidden messages are all recorded and scheduled for sending to the network so that the modulation can take place at the earliest possible time. The scheduler must consider how to balance the need to send the real online activities and the hidden messages. For instance, if there is no hidden message the real online activities may be held only for a period of time rather than be held forever. The system may need to allow human users to label real online activities about by when they must be posted, which will add further complexity to the scheduling algorithm.

While we have not investigated this extension in depth, we already identified two potential modulation methods. The first method considers the order of real online activities. Given a number of real online activities originated from a single human user, it is often possible to shuffle their order without causing any loss of information the human user intends to convey. By asking the human user to label dependent activities, the scheduler can know what activities are independent of each other so their relative positions in the sequence of finally posted online activities do not matter. The order can be used as a side channel to carry hidden information.

Another modulation method considers timing information. Even when there are dependencies among all real online activities in the queue, it is often the case that when exactly those activities go online does not really matter. This allows the possibility to use timing as a side channel to carry hidden information. Since it is normally impossible to control timing precisely, a mechanism is needed to calibrate the timing accuracy (possibly in a dynamic manner) and encode this information as part of the (expanded) hidden message.

Essentially speaking, modulating real online activities for information hiding is about using some non-content properties of real online activities as cover of the hidden information, so in theory it is still covered by the basic idea of the information hiding framework.

\subsection{Combination with Traditional Information Hiding Methods}
\label{section:CascadedIH}

It is possible to combine the proposed information hiding framework with one or more traditional content-based information hiding methods. The capacity of the proposed information hiding framework is smaller than most traditional methods since the latter can make use of a large amount of bits representing the contents of some digital objects. However, using traditional methods for transmitting short messages too frequently may invite suspicion (e.g., uploading images to a web forum too often may be seen as an abnormal behavior if image-based steganography is used). Combining the proposed information hiding framework and traditional ones can help increase the capacity of the proposed framework and reduce the unnecessary exposure of the use of traditional methods thus leading to a higher level of security.

The combination can be done by using the proposed information hiding framework as a command and control (C\&C) channel to signal the occasional use of one or more traditional content-based information hiding methods, the cover(s) and parameters used. Note that the covers of the traditional content-based information hiding methods do not have to be on the same web site or the Internet service as the proposed information hiding framework, which can help increase security.

\subsection{Naturalness Checking Sub-Algorithm}

While the naturalness checking sub-algorithm in the decoding process is not absolutely necessary, it can help increase capacity (by reducing the number of markers needed and their sizes) and improve usability of the system (by reducing unnecessary human involvement in the decoding process). The naturalness checking process normally requires a sufficiently long message so that there are enough linguistic or statistical evidence to make a decision. One typical example is the information entropy of the message decoded. A rubbish message normally looks random and tends to have a high information entropy, while a real textual message readable to human users tends to have a low information entropy. To calculate the information entropy of a message, there should be enough bytes to make a statistical difference. Some other statistical indicators can be developed based on a similar idea. For instance, for an English textual message we can look at the probability that all bytes received are either English letters or some commonly used punctuation markers to check its naturalness.

\section{Conclusion}

A new general framework for information hiding is proposed in this paper. The secret message is represented as activities happening in a selected digital world, conducted by one or more human and computer entities. The new framework relates to a number of old and new concepts in information hiding, but none of existing methods or frameworks cover all features offered by the new framework. We have developed a simple proof-of-concept system to demonstrate the feasibility of the framework. We expect more diverse and complicated implementations of the proposed framework can be done, and we are in the process of producing one involving multiple web services and with an enhanced level of security.

\section*{Acknowledgments}

Part of the work done by the first and second authors was funded by the Defence Science and Technology Laboratory (Dstl), UK through the Innovate UK's Small Business Research Initiative (SBRI) in 2014 (project title ``Mobile Magic Mirror (M3): Steganography and Cryptography on the move'').

\bibliographystyle{ACM-Reference-Format}
\balance
\bibliography{main}

\end{document}